\documentstyle[twoside,fleqn,espcrc2]{article}

\newcommand{\AmS}{{\protect\the\textfont2
  A\kern-.1667em\lower.5ex\hbox{M}\kern-.125emS}}

\title{  Critical parameters of $ N$-vector spin models on 3d lattices
 from high temperature series extended to order 
$ {  \beta^{21}}$ \thanks{ Presented by P. Butera}}
\author  { P. Butera  and M. Comi \address 
{INFN and Dept. of Physics, Milano University, 16 Via Celoria, 20133 
Milano(Italy)}}

\begin{document}

\begin{abstract}
High temperature expansions for the free energy, the  susceptibility and
the second correlation moment of the classical $N$-vector model 
[also denoted as the $O(N)$ symmetric classical spin Heisenberg  model or
as the lattice $O(N)$ nonlinear sigma model] 
 have been extended  to   order $\beta^{21}$ on the simple cubic 
 and the body centered cubic 
lattices, for arbitrary $N$.
The series for the second field derivative of the susceptibility has been
extended  to   order $\beta^{17}$.  An analysis of the newly computed series
yields  updated estimates 
of the model's critical parameters  in  good agreement 
with  present renormalization group estimates.
\end{abstract}
\maketitle
 \section{Introduction} 
We have extended to order $ {\beta^{21}}$ 
the computation of the High Temperature (HT) series for the 
free energy, the susceptibility $\chi(\beta,N)$, 
the second correlation moment $\mu_2(\beta,N)$ and
the second field derivative of the susceptibility 
$\chi_4(\beta,N)$ for the classical $N$-vector model
with any $N$  and on 
 all bipartite lattices, e.g. the (hyper) simple cubic (sc) and 
(hyper) body centered cubic (bcc). 
 The HT coefficients
 are written as explicit (rational) functions of $ N$ and it is also possible
 to exhibit their (polynomial) dependence on the  space dimensionality $d$.

The model Hamiltonian is 
$$
 H \{ v \} = -{1/2} \sum_{\langle \vec x,{\vec x}' \rangle }
v(\vec x) \cdot v({\vec x}')
$$

where $ v(\vec x)$ is a  $ N$-component classical unit spin.
As well known,  for $ N=0$ we get the   self-avoiding walk
 model, for  $ N=1$ the  Ising 
$ S= 1/2$ model, 
for $ N=2$ the XY model, for $ N=3$ the classical Heisenberg model,
for $ N \rightarrow \infty$ the (exactly solvable) spherical model.
We have used the vertex-renormalized
  Linked Cluster Expansion (LCE) method. 
Due to lack of space we  address the 
reader to  Refs.\cite{bc97,bc95,bc96,bc93} for an extensive
bibliography  on the LCE technique 
 and on the HT series published and analyzed
 before our work.
Here we shall only quote the paper 
by  M. L\"uscher and P.Weisz  \cite{lw88}
who computed
HT series by LCE  through  $ {\beta^{14}}$ for
 $ \chi$, $ \mu_{2}$ 
and    $ \chi_4$ on the sc lattices in $d=2,3,4$. 
 Our work has been made possible by an extensive redesign
of the algorithms proposed in Ref. \cite{lw88} in order to 
 reduce  drastically the  growth
of the computational weight with the order of the expansion. 
It is well known that the complexity of the HT expansions grows exponentially 
with the order  so that obtaining  one more series coefficient requires 
an effort 
substantially larger than for all previous ones.
Let us now sketch the
 calculation of $\chi$ through order $ {\beta^L}$.
 First we have to list all topologically  inequivalent multigraphs of 
the class $ S_{2}(L)$ defined as follows:
i) they have   2 external lines and no more than L internal lines; 
ii) they are  connected, 1-vertex, 1-line irreducible, and contain 
no odd loops;  
 iii) they have only  even degree vertices.
The Feynman rules are:
for each graph $ G  \in S_{2}(L) $ we have to     compute 
$$
 { F(G;N;d;\beta) = {I(G,d)C(G,N)W(G,N,d,\beta) \over S(G)}}
$$
where 
 $ S(G)$ is the  symmetry number of  the graph $G$
 (an integer number depending only on the  structure of $G$);  
 $ C(G,N)$ is the group coefficient of  $G$,  related to the $O(N)$ symmetry
   group  of the  model (a polynomial in $N$ with integer coefficients);
$ I(G,d)$ is an integer number counting the  unrestricted lattice per-site 
embeddings of $G$ (it depends  on  $G$ and the   lattice structure
 and  dimensionality $d$); 
$W(G,N,d,\beta)$ is associated to the vertices 
of $G$: it depends  
on the structure of the spin interaction and 
accounts ("renormalization") 
for all vertex insertions in the  graphs $ \in S_{2}(L)$
   (it is a  polynomial in $ {\beta}$ and $d$ and  
a rational function of $ N$).
Then we sum over all $F(G;N;d;\beta)$ 
$$
 {{ \chi(N,\beta)_{1LI} =   
 \sum_{G \in  S_2(L)}  F(G;N;d;\beta)}}
$$
 and finally
$$
 {{ \chi(N,\beta) ={  \chi(N,\beta)_{1LI} \over  1 - 4d\beta 
\chi(N,\beta)_{1LI} }}}
$$
 It is now clear that in order to get long  
series we need fast and efficient combinatorial 
algorithms for producing and comparing graphs, since
 the sum extends to all {\it topologically inequivalent} graphs  in   
$ S_{2}(L)$.
But generation algorithms inevitably produce duplicate 
graphs which  must be  discarded and
identification algorithms have a computational weight 
exponentially increasing with the number of vertices!
Fast algorithms are also needed to compute the symmetry number 
$ S(G)$ of each graph: this is also  a task of exponential complexity.
Moreover the computational load of the algorithm for the 
embedding number $I(G,d)$ grows 
fastly with $d$. 

The intricacies and the size  of this calculation  can be gauged from
 the  number of graphs $ {\approx 2\cdot 10^7}$,  
which enter into the computation 
of  $ { \chi}$  through order $ L=21$.
This figure should be compared with the corresponding 
one $ \approx 7 \cdot 10^3$ of the O($\beta^{14}$)  computation in 
Ref.\cite{lw88}.
However   we are still far from our present computational limits!

It is also worth to stress that  the LCE
 gives immediate access also to 
series for  the more general model  
described by the partition function: 
$$
  {  Z = \int 
 \Pi d\mu(\vec \varphi_i^2)  exp[ \beta \sum_{\langle i,j \rangle}\vec \varphi_i \cdot \vec 
\varphi_j]}
$$
 where $ \vec \varphi_i$ is a $ N$-component vector.
 By  a proper choice of the single spin measure $d\mu(\vec \varphi_i^2)$   
we obtain the HT series for a variety of models  
including  
 the general scalar isovector $P(\vec \varphi^2)$ lattice field 
theory, 
the general spin $S$ Ising model, the  Blume-Capel model,  
the double Gaussian model..., and this enables us to study various
representatives of each universality class.
The expected returns of our laborious enterprise include 
more accurate universality  tests, 
comparisons with estimates of  critical  exponents and
 of  critical amplitude ratios  obtained within
 the Renormalization Group (RG) approach\cite{zinn} either 
by the Fisher-Wilson perturbative
  $\epsilon$-expansion $(\epsilon =4-d)$ of
 the continuum $(\vec \varphi^2)^2$ model,
 or by the Parisi  coupling constant 
 expansion  of the same model in fixed dimension $d=3$, 
or else by other numerical methods. 
We also expect that the new feature of our calculation, namely
the explicit dependence on $N$ and $d$ of our HT coefficients, can provide
further insight into the properties of $1/N$ and $\epsilon $ expansions.

This project  has been carried on by an
 IBM Risc 6000/580 w-station 
with 128 Mbytes memory and  1.5 Gbytes  disk storage.
Typical cpu times are a few hours. 

\section {Series Analysis}
 
Our HT series  on the square lattice  have been tabulated and examined in 
Refs.\cite{bc96,bc93},
 while a study of the  3d case  has been partly presented in 
Refs.\cite{bc97,bc95}.
In  the 2d case, our results  can be summarized as follows: 
the HT series exhibit in the range $N \leq 2$ a qualitative behavior which
is sharply different from that for $N>2$.
For $N=2$ the critical properties of the model
 are completely  consistent with the  current ideas on 
the  Kosterlitz-Thouless phase transition.  For $N>2$  they appear to be 
consistent  
 with the conventional asymptotic freedom 
expectations from the perturbative RG  analysis
and with exact (although not completely rigorous) results from the Bethe
 Ansatz.
Similar conclusions  from an independent computation of 2d HT expansions
have been reached in Ref.\cite{campo}.
 In the 3d case,
 we have a  more conventional scenario: for all $ N$ a   second order
 phase transition occurs at nonzero temperature. In particular, 
if we set $ {\tau \equiv 1 - \beta/\beta_c}$, 
 in the vicinity of the critical point 
$ \chi$ and $ \xi$ behave as 
$$
 \chi(N,\beta) \simeq 
 A_{\chi}(N)\tau^{-\gamma(N)} (1+ a_{\chi}(N)
\tau^{\theta(N)} + 
$$
$$
a'_{\chi}(N) \tau^{2\theta(N)}+..+ e_{\chi}(N)\tau +.. )
$$
and 
$$
 \xi(N,\beta) \simeq 
A_{\xi}(N)\tau^{-\nu(N)} (1+ a_{\xi}(N)
\tau^{\theta(N)} + 
$$
$$
a'_{\xi}(N)\tau^{2\theta(N)}+..+ e_{\xi}(N)\tau +.. )
$$
when $\tau \downarrow 0$.   
The critical exponents $\gamma(N)$, $\nu(N)$ 
and the  scaling correction
exponent $ \theta(N)$ are universal (for each $ N$). 
On the contrary the critical  
amplitudes  $A_{\chi}(N), a_{\chi}(N),.., A_{\xi}(N), a_{\xi}(N)..$ 
 are nonuniversal, but amplitude ratios like $  {a_{\xi}(N)/a_{\chi}(N)}$ 
etc. are universal and interesting since they are not yet accurately known.

The numerical problem of determining simultaneously the critical parameters 
$\beta_c(N)$,$\gamma(N)$, $\nu(N)$  etc. as well as
 the  leading correction exponent $\theta(N)$ 
is  a difficult job: it amounts to an 
intrinsically  unstable double   exponential fit.
 However it must be faced, because
 without proper allowance at least for the main corrections to scaling 
no  improvement 
of the accuracy in  the determination of critical parameters can be warranted
even from  extended series.
We have therefore performed two kinds  of analysis  of the series 
 by the differential 
approximant\cite{gutt} method:

i) an unbiased analysis, where we do not assume to know some 
critical parameters when trying to determine the remaining ones;

ii) a biased analysis, where we   assume that the exponents 
of the  confluent corrections to scaling,  
take the values indicated by the fixed  
dimension RG perturbative calculations  in Ref.\cite{zinn} 
for $N \leq 3$ and in Ref.\cite{soko} for  $N > 3$.

From the analysis of  $ {\chi}$ and $  {\xi}$,
 we obtain in this way numerical estimates of  critical parameters which 
 are shown in two extensive tables of Ref.\cite{bc97}, 
but  cannot be reported here for lack of space. 
We can however summarize the 
qualitative conclusions.
As a first general result, both analyses confirm the traditional 
expectation that the bcc series have much better convergence 
properties than the sc series. 
The results of the unbiased series analysis
 agree well with 
 the exponent estimates from the
 fifth order   
$ \epsilon$-expansion
over the  range 
$0\leq N \leq 3 $ on which they are available.
 Analogously,  we observe an even closer agreement   with the
 results from the  fixed dimension coupling constant six loop
 expansion\cite{zinn,anto} over the whole range
 of values of $ N$,
and, for $0\leq N \leq 3 $, 
with the seven loop results\cite{mur}.
However for $N \geq 4$, the presence of residual trends in  
the extrapolations  of the  series  estimates for the exponents which use an 
 increasing number of  HT coefficients, seriously questions the 
accuracy of the unbiased analysis  suggesting  that  the confluent
 corrections are large and inadequately accounted for.
Therefore a biased analysis is necessary.

From the biased analysis we  also obtain
 interesting and, in some cases, accurate
 estimates of various critical amplitudes 
which confirm that scaling corrections
should not be neglected. 
For $N\leq3$, our biased exponent estimates  essentially agree with the
 unbiased ones, but  
they have a greater accuracy.
For $N\geq4$,  due to the large confluent corrections,  the biased estimates 
of the critical exponents differ up to a few percents both from the  unbiased 
ones and from the fixed dimension six loop perturbation results, 
suggesting that both a seven loop computation and  a more 
accurate evaluation of the renormalized coupling may be needed.


\begin{thebibliography}{9}

\bibitem{bc97} P.Butera and M.Comi, Phys. Rev. B56 (1997) 8212
(hep-lat/9703018).

\bibitem{bc95} P.Butera and M.Comi, Phys. Rev. B52 (1995) 6185.

\bibitem{bc96} P.Butera and M.Comi, Phys. Rev. B54 (1996) 15828.

\bibitem{bc93} P.Butera and M.Comi, Phys. Rev. B47 (1993) 11969.  

\bibitem{lw88} M. L\"uscher and P.Weisz,  Nucl. Phys. B300 (1988) 325.

\bibitem{zinn}  J. Zinn Justin, {\it Quantum field theory and critical
phenomena} (Clarendon, Oxford,1996).
\bibitem{campo} M. Campostrini,A. Pelissetto, P.Rossi, E. Vicari, Phys. Rev.  
D54 (1996) 1782.

\bibitem{gutt} A. J. Guttmann, in {\it Phase Transitions and critical Phenomena}, edited by
C. Domb and J. Lebowitz, (Academic, New York, 1989) Vol. 13.

\bibitem{soko} A. I. Sokolov, unpublished.

\bibitem{anto} S. A. Antonenko and A. I. Sokolov,  Phys. Rev. E51
(1995) 1894.

\bibitem{mur}  D. B. Murray and B. G. Nickel, Unpublished Guelph 
University report (1991).
\end{thebibliography}
\end{document}